\documentclass[twocolumn]{aastex63}

\accepted{2020}

\submitjournal{ApJL}

\shorttitle{DISCOVERY OF POLAR STREAM LMS-1}
\shortauthors{Yuan et al.}
\begin{document}

\title{A Low-Mass Stellar-Debris Stream Associated with a Globular Cluster Pair in the Halo}

\correspondingauthor{Zhen Yuan}
\email{sala.yuan@gmail.com}

\author{Zhen Yuan}
\affiliation{Key Laboratory for Research in Galaxies and Cosmology, Shanghai Astronomical Observatory, Chinese Academy of Sciences, 80 Nandan Road, Shanghai 200030, China}

\author{Jiang Chang}
\affiliation{Key Lab of Optical Astronomy, National Astronomical Observatories, CAS, 20A Datun Road, Chaoyang District, 100012 Beijing, China}
\affiliation{Purple Mountain Observatory, CAS, No.8 Yuanhua Road, Qixia District, Nanjing 210034, China}

\author{Timothy C. Beers}
\affiliation{Department of Physics and JINA Center for the Evolution of the Elements (JINA-CEE), University of Notre Dame, Notre Dame, IN 46556  USA}

\author{Yang Huang}
\affiliation{South-Western Institute For Astronomy Research, Yunnan University, Kunming 650500, P. R. China}

\begin{abstract}

There are expected to be physical relationships between the globular clusters (GCs) and stellar substructures in the Milky Way, not all of which have yet been found. We search for such substructures from a combined halo sample of SDSS blue horizontal-branch and SDSS+LAMOST RR Lyrae stars, cross-matched with astrometric information from $Gaia$ DR2. This is a sample of old stars which are also excellent tracers of structures, ideal for searching for ancient relics in the outer stellar halo. By applying the neural-network-based method $\textsc{StarGO}$ to the full 4D dynamical space of our sample, we rediscover the Sagittarius Stream, and find the debris of the $Gaia$-Enceladus-Sausage (GES) and the Sequoia events in the outer halo, as well as their linkages with several GCs. Most importantly, we find a new, low-mass, debris stream associated with a pair of GCs (NGC 5024 and NGC 5053), which we dub LMS-1. This stream has a very polar orbit, and occupies a region between 10 to 20 kpc from the Galactic center. NGC 5024 (M53), the more-massive of the associated GC pair, is very likely the nuclear star cluster of a now-disrupted dwarf galaxy progenitor, based on the results from N-body simulations.

\end{abstract}


\section{Introduction}
\label{sec:intro}

According to the $\Lambda$CDM cosmological model, the Milky Way (MW) has grown to its current size through mergers with numerous neighboring dwarf galaxies. Thanks to the advent of the $Gaia$ mission \citep{gaia18}, the stellar debris from relatively massive merger events such as the $Gaia$-Enceladus-Sausage \citep[GES;][]{belokurov18,haywood18,helmi18,myeong18b} and the Sequoia \citep[Seq;][]{myeong19} have been identified in the inner stellar halo. In the outer stellar halo, the full 6D panoramic portrait of the Sagittarius (Sgr) stream has been obtained for the first time \citep{antoja20,ibata20,ramos20}. These well-studied substructures and streams are the fossils from dwarf galaxies with dark matter halo masses of 10$^{10}$ -- 10$^{11}$M$_{\odot}$. The relics from less-massive dwarf galaxies engulfed by the MW are far more difficult to identify. However, based on the hierarchical paradigm of galaxy formation, the majority of the building blocks of the MW are expected to be small ($\la$ 10$^{9}$M$_{\odot}$). The identification of numerous minor mergers is thus essential for unraveling the complete assembly history of the MW.

Also importantly, low-mass dwarf galaxies have relatively short star-formation histories, and thus can provide direct records of the high-redshift ($z\sim$ 5) Universe; the epoch when globular clusters (GCs) were also formed. These clusters later fell into the MW as their host galaxies were disrupted, thus we would expect MW GCs to be connected to halo substructures. Indeed, such associations have been seen in M31 from recent photometric studies \citep[see, e.g.,][]{mackey10,huxor11,mackey19}. Similar relationships are less obvious in our Galaxy, as their detection relies on spectroscopic data for numerous individual stars. The few substructures in the MW known to be associated with GCs, the Sgr stream, the GES, and Sequoia, are all expected to have originated from massive accreted dwarf progenitors. Although GCs likely also populated less-massive progenitors, as has been found in nearby dwarf galaxies \citep[see, e.g.,][]{georgiev10}, their dwarf progenitors are expected to have been fully disrupted in the outer halo before sinking deep into the Galactic potential.

In order to identify the substructures associated with GCs stripped from lower-mass progenitors, we employ a sample of halo stars comprising two types of old stars, blue horizontal-branch (BHB) and RR Lyrae (RRL) stars. Such stars are not only representatives of the ancient halo, but are also excellent tracers of structure, owing to their precise distance estimates. Previous studies of substructure identification in dynamical space are limited to the inner-halo region, due to the lack of good distance estimates for more distant stars. The large range of distances of this halo sample, with an uncertainty as low as 5$\%$ (based on photometry only), allows us to identify dynamically tagged groups (DTGs) in the outer halo. In Sec.~\ref{sec:data}, we combine the SDSS BHB and SDSS+LAMOST RRL catalogs, and cross-match with $Gaia$ DR2 \citep{gaiadr2}, yielding $\sim$ 7600 stars with full 6D phase-space information. The group-identification approach is discussed in detail in Sec.~\ref{sec:method}. The DTGs with GC associations are presented in Sec.~\ref{sec:res}, including both existing and newly identified substructures. A summary and brief conclusions are provided in Sec.~\ref{sec:con}.

\section{Data}
\label{sec:data}
We combine a previous SDSS BHB catalog \citep{xue08} with the recently released SDSS+LAMOST RRL catalog \citep{liu20} to create a halo sample of 7640 stars that have full 6D kinematic information available. For BHB stars, line-of-sight velocities are derived from the SEGUE Stellar Parameter Pipeline \citep[SSPP;][]{lee08a,lee08b}, with uncertainties of 5 km s$^{-1}$ to 15 km s$^{-1}$ \citep{xue08}. The velocities of RRLs are taken from \citet{liu20}, which utilizes empirical templates to fit velocity curves of multiple measurements from the LAMOST \citep{deng12, zhao12} and the SDSS/SEGUE \citep{yanny09} surveys. Depending on the number of measurements, the velocity precision varies from 5 km s$^{-1}$ to 15 km s$^{-1}$. The distance estimates for both types of stars are obtained from multi-band photometry with mean uncertainties of about 5 $\%$ \citep{xue08, xue14,liu20}. We then cross-match the halo sample with $Gaia$ DR2 \citep{gaiadr2}. Given the magnitude range of the sample ($G\sim$ 17 -- 19), the errors of the proper motion measurements range from 0.13 to 0.60 mas yr$^{-1}$. Combining with the distance uncertainty of 5$\%$, the typical transverse velocity uncertainty is about 15 km s$^{-1}$ for the majority of stars in the sample, located 10 to 20 kpc from the Sun. This is equivalent to the uncertainty of the line-of-sight velocities. The resulting errors in the orbital parameters and other dynamical properties are sufficiently small to enable detection of significant groups in dynamical space. For the MW GCs, we employ the catalog from \citet{harris10}, with proper motions determined by \citet{eugene19}.

\section{Method}
\label{sec:method}

We apply the neural-network-based clustering method \textsc{StarGO} \citep{yuan18} to search for substructures that are clustered in the 4D space of orbital energy and angular momentum, both of which are approximately conserved, even in non-spherical potentials~\citep[e.g.,][]{helmi00}. The gravitational potential of \citet{mc17} is used to derive dynamical parameters with \textsc{AGAMA}~\citep{agama}. As in the previous work from \citet{yuan19,yuan20}, we use ($E$, $L$, $\theta$, $\phi$) as the input space, where the latter two angular parameters characterize directions of the orbital poles, and are defined as:
\begin{equation}
    \theta = \arccos(L_z/L), \qquad\qquad
    \phi = \arctan(L_x/L_y).
    \label{eq:angles}
\end{equation}
We use a 100$\times$100 neuron network, and follow a similar recipe as \citet{yuan20} to identify dynamical groups. Each grid point of the neuron map hosts one neuron, as shown in the left panel of Figure~\ref{fig:gi}, which has an initially randomized 4D weight vector. For a given input vector, the neuron that has the weight vector closest to it is defined as its best-matching unit (BMU). Each neuron updates its weight vectors to come closer to the input vector in the 4D space; the learning effectiveness depends on its distance to the BMU on the 2D map. The weight vectors of the neurons close to the BMU will be assimilated into the input vector more efficiently compared to those neurons located farther away. The final result of the learning process is a converged map, after a sufficient number of iterations. This process preserves the structure of the input data, and projects it onto a 2D map. 

The learning results can be revealed by differences in the weight vectors between adjacent neurons, which are defined as a 100$\times$100 $u$-matrix. In the left panel of Figure~\ref{fig:gi}, the gray-colored neurons have the top 20$\%$ $u$ values, denoting the lowest 20$\%$ similarities between neighbors. These neurons form gray boundaries, and separate the others into isolated islands (see the white patches in the left panel). Compared to the boundary neurons, those in islands have more similar weight vectors, and thus correspond to stars clustered in dynamical space. The idea of group identification is to find the islands isolated by the gray boundaries as we scan the threshold ($u_{\rm thr}$) that defines the boundary. We check the significance and contamination of neuron groups at each threshold value when they appear as isolated islands. By this means, we are able to systematically identify all the significant groups of stars clustered in the input space. In Figure~\ref{fig:gi}, we show all of the four groups identified in this work, for three different values of $u_{\rm thr}$. They are plotted by different colors on the neuron map shown in the left panel. The corresponding star groups form separate clusters in the input space shown in the right two panels.


Evaluation of the significance and contamination of each detected group is implemented as a post process. We first draw a Monte Carlo (MC) sample of 10,000 mock stars, based on the probability density function, in each dimension of the input space. Then we connect each mock star with its BMU on the trained neuron map, and obtain the probability ($p$) of a mock star being associated to a detected group $\mathcal{G}$ of $n$ members. The significance of $\mathcal{G}$ can be quantified by the binomial probability of detecting more than $n$ stars from the halo sample of $\mathcal{N}$ stars. The contamination can be derived as $p$/($n$/$\mathcal{N}$). We consider $\mathcal{G}$ as valid only if the significance is larger than 5 $\sigma$ and the contamination rate is less than 20$\%$. For each DTG, the valid members are re-verified by their probabilities ($p\geqslant$ 20$\%$) of being associated to the same group, after taking the observational uncertainties into account. The confidence of each DTG is derived as the average probability of its valid member stars being associated to it.

After validation of the detected groups, we check if any valid group is associated to known MW GCs. This is done by generating 1000 realizations for each GC, according to its observational uncertainties in 6D kinematics. As done for the mapping of mock stars, we connect each realization of a given GC with its BMU on the neuron map. The confidence level of the association between a GC and a DTG is quantified by the probability of the mock GC sample being associated with the same DTG. This value is used to compare the associations of different GCs with their DTGs.

\section{Results}
\label{sec:res}

\begin{figure*}
\centering
\includegraphics[width=\linewidth]{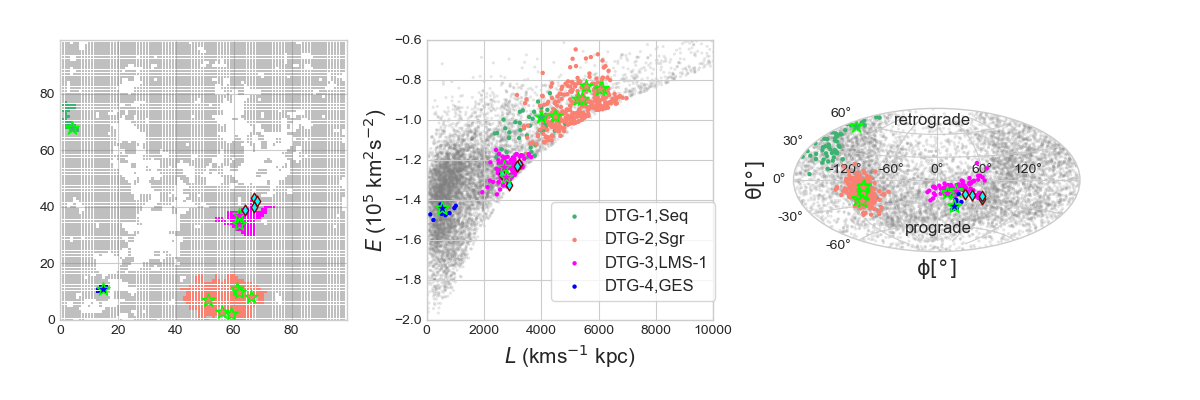}
\caption{The DTGs and associated GCs on the trained map (left panel), and in the input space ($E$, $L$, $\theta$, $\phi$) in the right two panels, where the latter two angles characterise the directions of orbital poles in Galactocentric coordinates. The gray circles in the right two panels represent the halo sample used in this study. The four DTGs are shown with different colors (green, salmon, magenta, blue), and form separated clumps in the input space. The GCs associated with each DTG are plotted by lime star symbols filled by the same color as the group color, and are well-within the corresponding clump. The four outliers stars are marked by cyan diamonds, which sit at the edge of DTG-3 on the neuron map and in the projection of orbital poles (see Sec.~\ref{subsec:ps} for details).}
\label{fig:gi}
\end{figure*}

\begin{figure*}
\centering
\includegraphics[width=\linewidth]{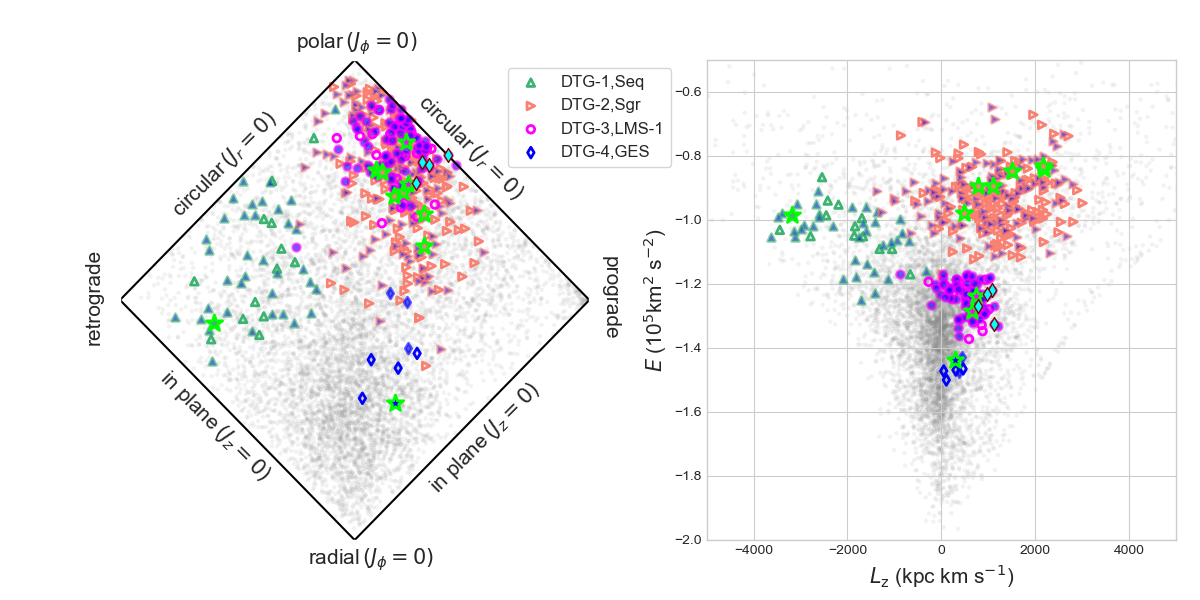}
\caption{The DTGs and their associated GCs in dynamical space. Left: The projected action-space map. The $x$-axis is ($J_{\phi}$/$J_{\rm{tot}}$), and the $y$-axis is ($J_{\rm z} - J_{\rm r}$)/$J_{\rm{tot}}$), where $J_{\rm{tot}}$ = $J_{\rm z} + J_{\rm r} + |J_{\phi}|$. Right: The space of orbital energy vs. the $z$-component of angular momentum. The gray circles represent the halo sample. The four DTGs are plotted by different colors and symbols, and the BHB members are filled by blue colors. The GCs are plotted by lime star symbols filled by the same color as the associated groups. DTG-1 (green upper triangles) has a very retrograde orbit, and DTG-4 (blue diamonds) exhibits prominent radial motion. Both DTG-2 (salmon right triangles) and DTG-3 (magenta circles) have fairly polar orbits, whereas the former has relatively higher energy.}
\label{fig:action}
\end{figure*} 

\begin{table*}[ht]
\centering
\caption{Properties of DTGs and Associated GCs}\label{tab:dtg}
\bgroup
\def\arraystretch{2}
	\begin{tabular}{|c|ccccccccc|}
		\hline
$u_{\rm{thr}}$ & $\mathcal{G}$ & $n_{\rm BHB}$ & $n_{\rm RRL}$ & $\mathcal{F}_{\mathrm{c}}$ &  Conf. & $\langle$[Fe/H]$\rangle$& $\sigma_{\rm [Fe/H]}$ (dex) & Substructure & Globular Clusters \\
 \hline
 $u_{1}$(60$^{\rm th}$)  & DTG-1     &29  &12 &14\% & 48\% &  $-$2.05$\pm$0.09  & 0.58$\pm$0.07 &Seq & NGC 6101 (100\%)\\
 $u_{2}$(45$^{\rm th}$)  & DTG-2    &116 &136  &6\%  &63\%  &  $-$1.87$\pm$0.02 & 0.31$\pm$0.02 &Sgr & Whiting 1 (78\%), M 54 (100\%), \\
 &&&&&&&&&  Terzan 7 (100\%), Arp 2 (99\%), \\
 &&&&&&&&&  Terzan 8 (100\%), Pal 12 (50\%)\\
 $u_{3}$(20$^{\rm th}$) & DTG-3   &75 &20  &9\%  &73\%  & $-$2.09$\pm$0.04  & 0.25$\pm$0.03 &LMS-1 & NGC 5024 (100\%), NGC 5053 (100\%)\\
 &DTG-4   &3 &5 &0\%  &39\% & $-$1.84$^{+0.24}_{-0.25}$ & 0.65$^{+0.21}_{-0.16}$  &GES  & NGC 6864 (M 75) (62\%)\\

\hline
\end{tabular}
\egroup
\end{table*}

In this work, we focus on the substructures that are dynamically associated with MW GCs. Although numerous valid DTGs could be identified from the trained neuron map, only those having strong associations with GCs are analyzed in this work. In total, we identify four DTGs at three different values of $u_{\rm thr}$. The details of these groups are summarized in Table~\ref{tab:dtg}, where $n_{\rm BHB}$ and $n_{\rm RRL}$ denote the number of group members from the SDSS BHB and SDSS+LAMOST RRL samples, respectively. The contamination fraction, $\mathcal{F}_{\rm c}$, and confidence level for each DTG are listed, as well as its mean and dispersion of [Fe/H]. The assigned substructures and associated GCs are shown in the last two columns. Since our halo sample is mainly populated by stars with [Fe/H] $\la-$1.5, all of the identified DTGs have very low mean metallicities ([Fe/H] $\approx$ $-$1.8 to $-$2.0). After taking into account the observational error of each group member star, the intrinsic dispersions of [Fe/H] of these DTGs are in the range of 0.25 -- 0.65 dex, which excludes the possibility of their progenitors being GCs. Figure~\ref{fig:gi} clearly shows that all the DTGs stand out from the gray background of the halo sample, and form separate clumps in the input space of ($E$, $L$) and ($\theta$, $\phi$). We note that, except for DTG-1 (green), which has a retrograde orbit with positive $\theta$, the other three groups, DTG-2 (salmon), DTG-3 (magenta), and DTG-4 (blue) all have prograde orbits with negative $\theta$. The GCs associated with each DTG are embedded well-within its individual clump, shown as lime star symbols filled by the same colors as their DTGs. The confidence level of each association, derived in the same way as that of each member star, is provided in parentheses following the GC name in the last column of Table~\ref{tab:dtg}.

Figure~\ref{fig:action} shows the location of the DTGs and their associated GCs in different dynamical-space visualizations, color-coded as in Figure~\ref{fig:gi}. We differentiate the BHB members from the RRL members by filling the former with blue colors. The left panel of Figure~\ref{fig:action} shows the projected action-space map. DTG-1 clearly occupies the corner of retrograde orbits, and DTG-4 is situated in the region representing radial orbits. DTG-2 and DTG-3 have fairly polar orbits, and significantly overlap with each other in this projection; note that DTG-2 has higher orbital energy than DTG-3, which makes them clearly separable in the ($E$, $L_{\rm z}$) space shown in the right panel. DTG-1 has slightly lower energy than DTG-2, but they have distinguishable distributions of $L_{\rm z}$. Among all the groups, DTG-4 has the lowest energy, as well as the lowest rotational motion. Utilizing these features of orbital properties, we analyze and assign an origin to each group below.

\subsection{Existing Substructures}
\label{subsec:exs}

The first valid group with a GC association is DTG-1, identified at $u_{\rm thr}$ = $u_{60\%}$. This is the only retrograde group found in this work, which consists of 29 BHB and 12 RRL stars. A great number of streams and substructures with retrograde motions have been reported by several studies \citep{grillmair06, malhan18, price18,myeong18a, malhan19, matsuno19, koppelman19, yuan20}. These groups are contributed by at least one substantial merger event that took place at an early epoch, Sequoia \citep{myeong19}, which also brought in several retrograde GCs. The distribution of DTG-1 in the action-space map and ($E$, $L_{\rm z}$) overlaps perfectly with that of the Sequoia groups found in the inner halo with $d\la$ 10 kpc \citep{matsuno19, myeong19, yuan20}. The Stars in DTG-1 reside at 10 kpc to 40 kpc from the Sun, but most of them have pericenter distances $\la$ 10 kpc, similar to the Sequoia relic, which is also consistent with their relatively low orbital energy $E\sim$ $-$10$^5$ to $-$1.2$\times$10$^5$ km$^2$ s$^{-2}$. This implies that DTG-1 comes from an early accreted dwarf galaxy, possibly the same one as Sequoia. DTG-1 is very likely part of the Sequoia debris that currently occupies the outer halo.

\begin{figure}
\centering
\includegraphics[width=\linewidth]{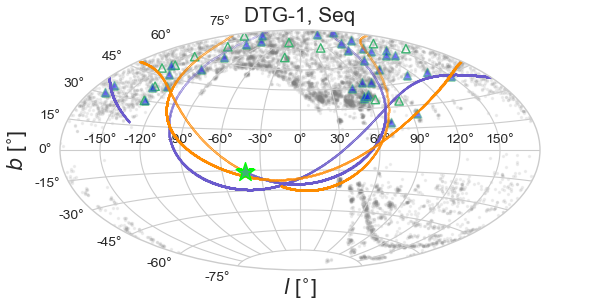}
\includegraphics[width=\linewidth]{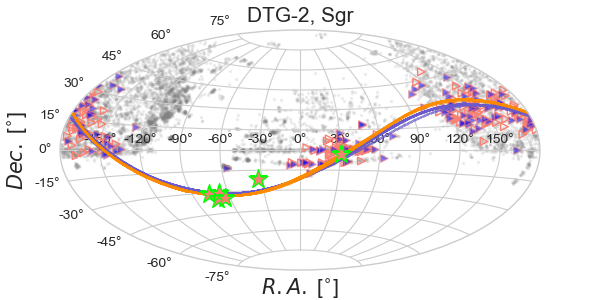}
\includegraphics[width=\linewidth]{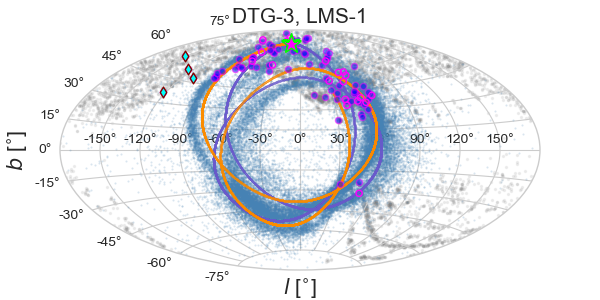}
\includegraphics[width=\linewidth]{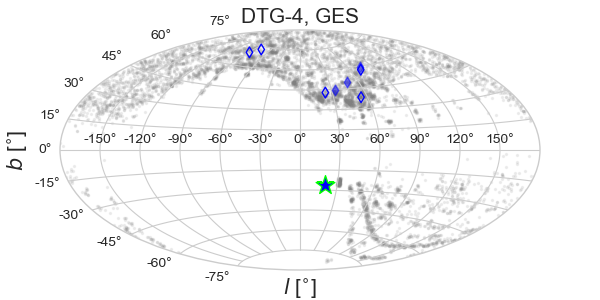}

\caption{The on-sky projection of four DTGs and their associated GCs, superposed on the halo sample, plotted in the same way as Fig.~\ref{fig:action}. First Panel: DTG-1 (Seq; green upper triangles) and NGC 6101 in Galactic coordinates. The forward-integrated orbit (purple) and backward orbit (orange) of the GC traverse the group members, which are all populated in the North. Second Panel: DTG-2 (the Sgr stream; salmon right triangles), in equatorial coordinates, and the six associated GCs, Whiting 1 (R.A. $\approx$ 30 $^{\circ}$), NGC 6715 (M 54), Terzan 7, Arp 2, Terzan 8, and Pal 12. Third Panel: DTG-3 (LMS-1; magenta circles) and its associated pair of GCs, in Galactic coordinates. The steel-blue circles denote the simulated stream on the orbit of NGC 5024 (purple: forward; orange: backward) with eight apocentric passages, coincident with the stream members in both hemispheres. The four outliers at $l\sim-$ 120$^{\circ}$ are highlighted by cyan diamonds. Fourth Panel: DTG-4 (GES; blue diamonds) and NGC 6864, in Galactic coordinates, which are located in the region of the VOD and HAC. }
\label{fig:stream}
\end{figure}

By applying the approach discussed in Sec.~\ref{sec:method}, we find that DTG-1 is dynamically associated with NGC 6101, which is also categorized as a Sequoia GC according to two different studies \citep{massari19,myeong19}. NGC 6101 has a mass of $\sim$ 10$^5$M$_{\odot}$ and low metallicity, [Fe/H] = $-$1.98 \citep{harris10}, consistent with the picture that it was born in the early star-formation epochs of a classical dwarf galaxy, and accreted to the MW during the merger. We integrate the orbit of NGC 6101 in forward and backward directions for about three orbital periods (1.5 Gyr, T$_{\mathrm{orb}}\approx$ 0.5 Gyr), shown in the first panel of Figure~\ref{fig:stream}. Although the GC is currently situated in the South, its orbit comes across the Galactic plane and traverses most of the stellar members above the plane.

The largest group among the four is DTG-2, identified at $u$ = $u_{45\%}$, which has 116 BHB and 136 RRL members, covering a large heliocentric distance range from 5 to 50 kpc. The on-sky projection of DTG-2 in equatorial coordinates reveals it as the Sgr stream (see the second panel of Figure~\ref{fig:stream}). We find that six GCs (Whiting 1, M 54, Terzan 7, Arp 2, Terzan 8, and Pal 12) are associated to DTG-2, all of which have been confirmed to be associated with the Sgr stream by previous studies \citep[see, e.g.,][]{law10,sohn18,bellazzini20}. Note that the association of Pal 12 has the lowest confidence level (50\%), among all the GC associations identified in this work.

Another group with distinctive orbital features is DTG-4, comprising three BHB member and five RRL stars, located between 7 to 15 kpc from the Galactic center, with a prominent radial motion and high orbital eccentricity, $e\sim$ 0.7. The orbital energy is fairly low, $E\sim$ $-$1.45$\times$10$^5$ km$^2$ s$^{-2}$, characterizing it as an inner-halo substructure. All of these properties suggest that DTG-4 very likely comes from the GES, which is an early, massive radial-merger event \citep{belokurov18,helmi18}. Its associated GC, M75 (NGC 6864), is also identified as a GES globular cluster by both \citet{myeong18b} and \citet{massari19}. We plot DTG-4 and its associated GC, in Galactic coordinates, in the bottom panel of Figure~\ref{fig:stream}. Six members at $l\sim$ 30$^{\circ}$ -- 60$^{\circ}$ are located in the region of the Northern Hercules Aquila Cloud (HAC), and NGC 6864 is in the Southern HAC \citep[][]{belokurov07,simion14}. The other two member at $l\approx-$60$^{\circ}$ are in the area of the Virgo Over-Density (VOD), which has been shown to share the same origin as the HAC \citep[see][and references]{simion19}.

\subsection{The Polar Stream LMS-1}
\label{subsec:ps}

There is only one group, DTG-3, that cannot be assigned to any existing substructures. DTG-3 has 75 BHB and 20 RRL members, shown as the magenta group in Figures~\ref{fig:gi} and \ref{fig:action}. It has intermediate orbital energy ($E\sim$ $-$1.4$\times$10$^5$ to $-$1.2$\times$10$^5$ km$^2$ s$^{-2}$), between DTG-4, representing the GES debris in the inner halo, and DTG-2, confirmed as the Sgr Stream in the outer halo. The stellar members of DTG-3 have Galactocentric distances $r\sim$ 10 -- 25 kpc and pericentric distances $r_{\rm p}\la$ 15 kpc. As for the Sgr stream identified in this work, its stellar members have $r\sim$ 15 -- 45 kpc, and $r_{\rm p}\la$ 30 kpc. This indicates that DTG-3 is situated closer to the Galactic center than the Sgr stream, consistent with its lower orbital energy. DTG-3 has an average orbital inclination angle of 80$^{\circ}$, which is more polar than the Sgr Stream (76$^{\circ}$) identified in this work. We plot the on-sky projection of DTG-3, in Galactic coordinates, shown as the magenta group in the third panel of Figure~\ref{fig:stream}. It spans a wide region on the sky, with a coverage of more than 100$^{\circ}$ in $l$, while maintaining a relatively coherent structure, like the Sgr stream shown in the second panel. Most of the stream members are populated in the North, with three members found in the South, owing to the limited sky coverage of both the SDSS and LAMOST surveys. We name this substructure as the low-mass stellar-debris stream (LMS-1), because it is a wide debris-stream similar to the Sgr stream, but is made up of much fewer members.

The two associated GCs are not only embedded in LMS-1 in dynamical space (see Figures~\ref{fig:gi} and \ref{fig:action}), but also in configuration space (see the on-sky projection in Figure~\ref{fig:stream}). The two GCs are currently located in the distance range of the LMS-1 members, at $r\approx$ 18 kpc, close to their apocenters ($r_{\rm a}\approx$ 20 kpc). They have pericentric distances $r_{\rm p} \approx$ 10 kpc, similar to the LMS-1 members. The separation between these two GCs is 500 pc, but their velocity difference is $\sim$ 200 km s$^{-1}$, which makes it a unique and intriguing pair. NGC 5024 (M53), in the rank of massive GCs in the MW, has a mass of $\sim$ 5$\times$10$^5$M$_{\odot}$ \citep{harris10}, and [Fe/H] = $-$2.07 \citep{boberg16}, which is much more massive than its companion NGC 5053, with a mass of $\sim$ 5$\times$10$^4$M$_{\odot}$ \citep{baumgardt17}, and [Fe/H] = $-$2.45 \citep{boberg15}. We emphasize that, although the pair of GCs are very close together, they are not bound to each other. However, it is very unlikely that they just happen to be passing by one another. \citet{chun10} showed that this GC pair is surrounded by a complex stellar envelope, using deep photometric data from MegaCam. \citet{ngeow20} claimed there are no extra-tidal RR Lyraes associated with them within $\sim$ 8 deg$^2$. \citet{massari19} attributed these two GCs to the Helmi Stream. In this work, we show that this GC pair is embedded in a wide stream, suggesting they were stripped from the same parent dwarf galaxy. This also naturally leads to the plausible scenario that the relatively massive GC (NGC 5024) could be the core of the dwarf galaxy progenitor of both the stellar stream and the GCs. To verify this possibility, we trace the orbit of NGC 5024 for about three periods of time ($T_{\rm orb}\approx$ 0.32 Gyr) in both backward (orange line) and forward (blue line) directions, as shown in Figure~\ref{fig:stream}. The trajectory of NGC 5024 traverses the majority of the stream members located in the North, with four outliers at $l\approx-$120$^{\circ}$ (cyan diamonds), and perfectly matches with the three Southern members as well. These outlier stars, sitting at the edge of the distribution of the orbital poles of LMS-1 (see the right panel of Figure~\ref{fig:gi}), are possibly the result of orbital precession. Although the orbit of NGC 5053 is very similar to NGC 5024, given their large mass difference, the nuclear star cluster of the dwarf galaxy progenitor of LMS-1 was very likely NGC 5024, whereas NGC 5053 was off-center, and stripped at a different epoch. This could explain why the two GCs are coincidentally almost at the same place now, but not in a dynamically bound system. When this paper was being reviewed, \citet{naidu20} reported the discovery of a new substructure, ''Wukong'', which has similar dynamical properties as LMS-1. There are three GCs (NGC 5024, NGC 5053, and ESO 280-SC06) attributed to ``Wukong'' in their studies, whereas only the first two is identified as dynamically associated to LMS-1 in this work.

\begin{figure*}
\centering
\includegraphics[width=\linewidth]{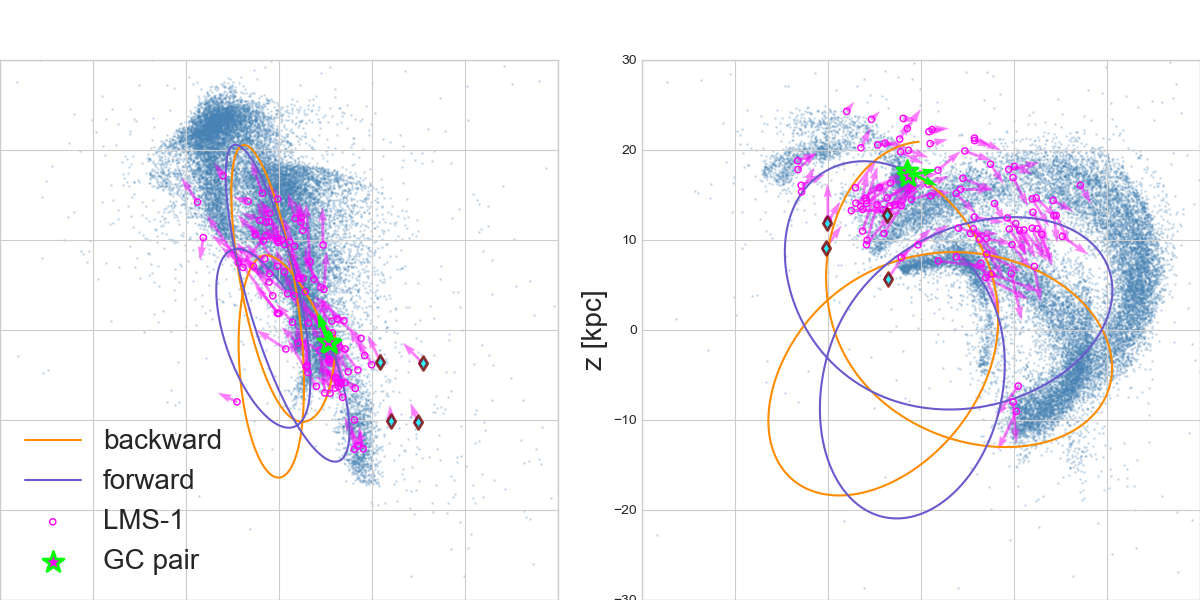}
\caption{LMS-1, with the pair of GCs, and the simulated stream in the observable sky (Dec. $>$ -10$^{\circ}$) on the orbit of NGC 5024 with eight apocentric passages, in Galactocentric coordinates ($x$, $y$) and ($y$, $z$), plotted in the same way as Figure~\ref{fig:stream}. The arrow of each stream member indicates its velocity vector, which generally follows the trajectory of the stream at its location. The observed members are very likely the mixture of stellar debris from multiple wraps stripped at different epochs.}
\label{fig:xyz}
\end{figure*}

All the DTGs identified in this work are fairly metal-poor, because the halo sample used is a combination of two types of old stars. However, it is noteworthy that the mean metallicity of LMS-1 ($\langle$[Fe/H]$\rangle$ = $-$2.09) is similar to that of the Seq DTG ($\langle$[Fe/H]$\rangle$ = $-$2.05), and is slightly lower than those of the Sgr stream and the GES groups ($\langle$[Fe/H]$\rangle\approx-$1.8), implying that the progenitor dwarf galaxies of both LMS-1 and Sequoia are less massive than the other two. The number of the LMS-1 members is about one-third of the number of the Sgr members in our sample of BHB and RR Lyrae stars. This is in-line with the number of GCs associated with these two streams: the Sgr stream has six GCs and LMS-1 has two. This again indicates that the dwarf galaxy progenitor of LMS-1 is smaller than the Sgr dwarf, but still massive enough to host two GCs. This might be the reason that LMS-1 was not discovered by previous photometric studies, as it has much lower surface brightness compared with the Sgr stream. We have simulated LMS-1 using a dwarf satellite with a total mass of 2$\times$10$^9$M$\odot$, given that the Sgr progenitor mass is $\sim$10$^{10}$M$\odot$. The initial condition of the satellite is derived by rewinding the orbit of NGC 5024 for eight orbital periods. The orbital information at its first apocentric passage is recorded as the initial condition of the dwarf satellite. We then let it evolve in the MW potential for about eight orbital periods ($\sim$2.4 Gyr). It is fully disrupted now, and has strewn stream members across the Galactic plane, shown as the blue scatter in the third panel of Figure~\ref{fig:stream}. As can be seen, the simulated stream agrees well with the integrated orbit of NGC 5024, and can fully cover the footprint of LMS-1, as also shown, in Galactic coordinates, in Figure~\ref{fig:xyz}.

\section{Conclusions}
\label{sec:con}

In this work, we employ two types of old stars, BHBs and RRLs, to construct a fair sample of $\sim$7600 stars in the ancient halo. Both are excellent distance indicators, thus we are able to obtain accurate 6D kinematic information for this halo sample, and derive their dynamical parameters. We apply the neural-network-based clustering method $\textsc{StarGO}$ to this sample in the space of orbital energy and angular momentum, and identify four DTGs that are confidently associated with known MW GCs. The largest group, DTG-2, is confirmed to be the Sgr Stream, and the other two, DTG-1 and DTG-4, are very likely the debris of Sequoia and GES left in the outer halo. 

We show that DTG-3 is a new stream, having a very polar orbit with inclination angle of 80$^{\circ}$, which we refer to as LMS-1. It is associated with a pair of GCs (NGC 5024 and NGC 5053), which is the first example of a low-mass stellar-debris stream with embedded GCs from a disrupted low-mass dwarf galaxy. By tracing the orbit of the more-massive member of the pair, NGC 5024, we suggest that it is probably the core of the dwarf galaxy progenitor of LMS-1 and NGC 5053. We then use the orbital information of NGC 5024 as the initial condition for a dwarf satellite with a total mass of 2$\times$10$^9$M$_{\odot}$, and run a N-body simulation in an analytic MW potential. The resulting stream nicely covers the observed LMS-1 in both hemispheres, with a few outliers possibly due to orbital precession.

The stellar system of LMS-1 and its pair of GCs belongs to the vast polar structure \citep[VPOS;][]{riley20}, initially suggested by \citet{pawlowski12}. Although recent studies have shown that the orbital poles of the existing stellar streams and GCs do not cluster around the direction of the VPOS, the LMS-1 system adds a substantial accreted dwarf to it. We believe that numerous additional low-mass stellar-debris streams remain to be discovered that could be associated with GCs in the outer halo. This will help build a more complete MW assembly history, as well as open a new window to study the formation and evolution of GCs in ancient dwarf galaxies.

\section*{Acknowledgements}
This work has made use of data from the European Space Agency (ESA) mission
{\it Gaia} (\url{https://www.cosmos.esa.int/gaia}), processed by the {\it Gaia}
Data Processing and Analysis Consortium (DPAC, \url{https://www.cosmos.esa.int/web/gaia/dpac/consortium}). Funding for the DPAC has been provided by national institutions, in particular the institutions
participating in the {\it Gaia} Multilateral Agreement.

The authors would like to thank the referee for constructive comments. Z.Y. wishes to thank Eugene Vasiliev for insightful discussions about the GC pair, and Iulia Simion for sharing her expertise on VOD and HAC. Z.Y. acknowledges the support from the LAMOST Fellow project, and Shanghai Sailing Program (Y955051001). J.C. is supported by Young Scientists Foundation of JiangSu Province (BK20181110), the NSFC fundings (No. 11825303, 11861131006) and the Alibaba Cloud Fellowship. T.C.B. acknowledges partial support from grant PHY 14-30152, Physics Frontier Center/JINA Center for the Evolution of the Elements (JINA-CEE), awarded by the US National Science Foundation. T.C.B. is also grateful for support from a PIFI Distinguished Scientist award from the Chinese Academy of Sciences. Y.H. is supported by National Natural Science Foundation of China grants 11903027, 11973001, 11833006, 11811530289, U1731108, and U1531244, and National Key R\&D Program of China No. 2019YFA0405503. 

\bibliography{ms}{}

\begin{thebibliography}{}
\expandafter\ifx\csname natexlab\endcsname\relax\def\natexlab#1{#1}\fi
\providecommand{\url}[1]{\href{#1}{#1}}
\providecommand{\dodoi}[1]{doi:~\href{http://doi.org/#1}{\nolinkurl{#1}}}
\providecommand{\doeprint}[1]{\href{http://ascl.net/#1}{\nolinkurl{http://ascl.net/#1}}}
\providecommand{\doarXiv}[1]{\href{https://arxiv.org/abs/#1}{\nolinkurl{https://arxiv.org/abs/#1}}}

\bibitem[{{Antoja} {et~al.}(2020){Antoja}, {Ramos}, {Mateu}, {Helmi}, {Anders},
  {Jordi}, \& {Carballo-Bello}}]{antoja20}
{Antoja}, T., {Ramos}, P., {Mateu}, C., {et~al.} 2020, \aap, 635, L3,
  \dodoi{10.1051/0004-6361/201937145}

\bibitem[{{Baumgardt}(2017)}]{baumgardt17}
{Baumgardt}, H. 2017, \mnras, 464, 2174, \dodoi{10.1093/mnras/stw2488}

\bibitem[{{Bellazzini} {et~al.}(2020){Bellazzini}, {Ibata}, {Malhan}, {Martin},
  {Famaey}, \& {Thomas}}]{bellazzini20}
{Bellazzini}, M., {Ibata}, R., {Malhan}, K., {et~al.} 2020, \aap, 636, A107,
  \dodoi{10.1051/0004-6361/202037621}

\bibitem[{{Belokurov} {et~al.}(2018){Belokurov}, {Erkal}, {Evans}, {Koposov},
  \& {Deason}}]{belokurov18}
{Belokurov}, V., {Erkal}, D., {Evans}, N.~W., {Koposov}, S.~E., \& {Deason},
  A.~J. 2018, \mnras, 478, 611, \dodoi{10.1093/mnras/sty982}

\bibitem[{{Belokurov} {et~al.}(2007){Belokurov}, {Evans}, {Bell}, {Irwin},
  {Hewett}, {Koposov}, {Rockosi}, {Gilmore}, {Zucker}, {Fellhauer},
  {Wilkinson}, {Bramich}, {Vidrih}, {Rix}, {Beers}, {Schneider}, {Barentine},
  {Brewington}, {Brinkmann}, {Harvanek}, {Krzesinski}, {Long}, {Pan},
  {Snedden}, {Malanushenko}, \& {Malanushenko}}]{belokurov07}
{Belokurov}, V., {Evans}, N.~W., {Bell}, E.~F., {et~al.} 2007, \apjl, 657, L89,
  \dodoi{10.1086/513144}

\bibitem[{{Boberg} {et~al.}(2015){Boberg}, {Friel}, \& {Vesperini}}]{boberg15}
{Boberg}, O.~M., {Friel}, E.~D., \& {Vesperini}, E. 2015, \apj, 804, 109,
  \dodoi{10.1088/0004-637X/804/2/109}

\bibitem[{{Boberg} {et~al.}(2016){Boberg}, {Friel}, \& {Vesperini}}]{boberg16}
---. 2016, \apj, 824, 5, \dodoi{10.3847/0004-637X/824/1/5}

\bibitem[{{Chun} {et~al.}(2010){Chun}, {Kim}, {Sohn}, {Park}, {Han}, {Kim},
  {Lee}, {Lee}, {Lee}, \& {Sohn}}]{chun10}
{Chun}, S.-H., {Kim}, J.-W., {Sohn}, S.~T., {et~al.} 2010, \aj, 139, 606,
  \dodoi{10.1088/0004-6256/139/2/606}

\bibitem[{{Deng} {et~al.}(2012){Deng}, {Newberg}, {Liu}, {Carlin}, {Beers},
  {Chen}, {Chen}, {Christlieb}, {Grillmair}, {Guhathakurta}, {Han}, {Hou},
  {Lee}, {L{\'e}pine}, {Li}, {Liu}, {Pan}, {Sellwood}, {Wang}, {Wang}, {Yang},
  {Yanny}, {Zhang}, {Zhang}, {Zheng}, \& {Zhu}}]{deng12}
{Deng}, L.-C., {Newberg}, H.~J., {Liu}, C., {et~al.} 2012, Research in
  Astronomy and Astrophysics, 12, 735, \dodoi{10.1088/1674-4527/12/7/003}

\bibitem[{{Gaia Collaboration} {et~al.}(2018){Gaia Collaboration}, {Brown},
  {Vallenari}, {Prusti}, {de Bruijne}, {Babusiaux}, {Bailer-Jones}, {Biermann},
  {Evans}, {Eyer}, \& et~al.}]{gaia18}
{Gaia Collaboration}, {Brown}, A.~G.~A., {Vallenari}, A., {et~al.} 2018, \aap,
  616, A1, \dodoi{10.1051/0004-6361/201833051}

\bibitem[{{Georgiev} {et~al.}(2010){Georgiev}, {Puzia}, {Goudfrooij}, \&
  {Hilker}}]{georgiev10}
{Georgiev}, I.~Y., {Puzia}, T.~H., {Goudfrooij}, P., \& {Hilker}, M. 2010,
  \mnras, 406, 1967, \dodoi{10.1111/j.1365-2966.2010.16802.x}

\bibitem[{{Grillmair} \& {Dionatos}(2006)}]{grillmair06}
{Grillmair}, C.~J., \& {Dionatos}, O. 2006, \apjl, 643, L17,
  \dodoi{10.1086/505111}

\bibitem[{{Harris}(2010)}]{harris10}
{Harris}, W.~E. 2010, ArXiv e-prints.
\newblock \doarXiv{1012.3224}

\bibitem[{{Haywood} {et~al.}(2018){Haywood}, {Di Matteo}, {Lehnert}, {Snaith},
  {Khoperskov}, \& {G{\'o}mez}}]{haywood18}
{Haywood}, M., {Di Matteo}, P., {Lehnert}, M.~D., {et~al.} 2018, \apj, 863,
  113, \dodoi{10.3847/1538-4357/aad235}

\bibitem[{{Helmi} {et~al.}(2018){Helmi}, {Babusiaux}, {Koppelman}, {Massari},
  {Veljanoski}, \& {Brown}}]{helmi18}
{Helmi}, A., {Babusiaux}, C., {Koppelman}, H.~H., {et~al.} 2018, \nat, 563, 85,
  \dodoi{10.1038/s41586-018-0625-x}

\bibitem[{{Helmi} \& {de Zeeuw}(2000)}]{helmi00}
{Helmi}, A., \& {de Zeeuw}, P.~T. 2000, \mnras, 319, 657,
  \dodoi{10.1046/j.1365-8711.2000.03895.x}

\bibitem[{{Huxor} {et~al.}(2011){Huxor}, {Ferguson}, {Tanvir}, {Irwin},
  {Mackey}, {Ibata}, {Bridges}, {Chapman}, \& {Lewis}}]{huxor11}
{Huxor}, A.~P., {Ferguson}, A.~M.~N., {Tanvir}, N.~R., {et~al.} 2011, \mnras,
  414, 770, \dodoi{10.1111/j.1365-2966.2011.18450.x}

\bibitem[{{Ibata} {et~al.}(2020){Ibata}, {Bellazzini}, {Thomas}, {Malhan},
  {Martin}, {Famaey}, \& {Siebert}}]{ibata20}
{Ibata}, R., {Bellazzini}, M., {Thomas}, G., {et~al.} 2020, \apjl, 891, L19,
  \dodoi{10.3847/2041-8213/ab77c7}

\bibitem[{{Koppelman} {et~al.}(2019){Koppelman}, {Helmi}, {Massari},
  {Price-Whelan}, \& {Starkenburg}}]{koppelman19}
{Koppelman}, H.~H., {Helmi}, A., {Massari}, D., {Price-Whelan}, A.~M., \&
  {Starkenburg}, T.~K. 2019, \aap, 631, L9, \dodoi{10.1051/0004-6361/201936738}

\bibitem[{{Law} \& {Majewski}(2010)}]{law10}
{Law}, D.~R., \& {Majewski}, S.~R. 2010, \apj, 718, 1128,
  \dodoi{10.1088/0004-637X/718/2/1128}

\bibitem[{{Lee} {et~al.}(2008{\natexlab{a}}){Lee}, {Beers}, {Sivarani},
  {Allende Prieto}, {Koesterke}, {Wilhelm}, {Re Fiorentin}, {Bailer-Jones},
  {Norris}, {Rockosi}, {Yanny}, {Newberg}, {Covey}, {Zhang}, \& {Luo}}]{lee08a}
{Lee}, Y.~S., {Beers}, T.~C., {Sivarani}, T., {et~al.} 2008{\natexlab{a}}, \aj,
  136, 2022, \dodoi{10.1088/0004-6256/136/5/2022}

\bibitem[{{Lee} {et~al.}(2008{\natexlab{b}}){Lee}, {Beers}, {Sivarani},
  {Johnson}, {An}, {Wilhelm}, {Allende Prieto}, {Koesterke}, {Re Fiorentin},
  {Bailer-Jones}, {Norris}, {Yanny}, {Rockosi}, {Newberg}, {Cudworth}, \&
  {Pan}}]{lee08b}
---. 2008{\natexlab{b}}, \aj, 136, 2050, \dodoi{10.1088/0004-6256/136/5/2050}

\bibitem[{{Lindegren} {et~al.}(2018){Lindegren}, {Hern{\'a}ndez}, {Bombrun},
  {Klioner}, {Bastian}, {Ramos-Lerate}, \& {de Torres}}]{gaiadr2}
{Lindegren}, L., {Hern{\'a}ndez}, J., {Bombrun}, A., {et~al.} 2018, \aap, 616,
  A2, \dodoi{10.1051/0004-6361/201832727}

\bibitem[{{Liu} {et~al.}(2020){Liu}, {Huang}, {Zhang}, {Xiang}, {-J.}, {Ren},
  {Chen}, {Yuan}, {Wang}, {Yang}, {Tian}, {Wang}, \& {Liu}}]{liu20}
{Liu}, G.~C., {Huang}, Y., {Zhang}, H.~W., {et~al.} 2020, arXiv e-prints,
  arXiv:2002.01188.
\newblock \doarXiv{2002.01188}

\bibitem[{{Mackey} {et~al.}(2010){Mackey}, {Huxor}, {Ferguson}, {Irwin},
  {Tanvir}, {McConnachie}, {Ibata}, {Chapman}, \& {Lewis}}]{mackey10}
{Mackey}, A.~D., {Huxor}, A.~P., {Ferguson}, A.~M.~N., {et~al.} 2010, \apjl,
  717, L11, \dodoi{10.1088/2041-8205/717/1/L11}

\bibitem[{{Mackey} {et~al.}(2019){Mackey}, {Ferguson}, {Huxor}, {Veljanoski},
  {Lewis}, {McConnachie}, {Martin}, {Ibata}, {Irwin}, {C{\^o}t{\'e}},
  {Collins}, {Tanvir}, \& {Bate}}]{mackey19}
{Mackey}, A.~D., {Ferguson}, A.~M.~N., {Huxor}, A.~P., {et~al.} 2019, \mnras,
  484, 1756, \dodoi{10.1093/mnras/stz072}

\bibitem[{{Malhan} {et~al.}(2019){Malhan}, {Ibata}, {Carlberg}, {Bellazzini},
  {Famaey}, \& {Martin}}]{malhan19}
{Malhan}, K., {Ibata}, R.~A., {Carlberg}, R.~G., {et~al.} 2019, \apjl, 886, L7,
  \dodoi{10.3847/2041-8213/ab530e}

\bibitem[{{Malhan} {et~al.}(2018){Malhan}, {Ibata}, \& {Martin}}]{malhan18}
{Malhan}, K., {Ibata}, R.~A., \& {Martin}, N.~F. 2018, \mnras, 481, 3442,
  \dodoi{10.1093/mnras/sty2474}

\bibitem[{{Massari} {et~al.}(2019){Massari}, {Koppelman}, \&
  {Helmi}}]{massari19}
{Massari}, D., {Koppelman}, H.~H., \& {Helmi}, A. 2019, \aap, 630, L4,
  \dodoi{10.1051/0004-6361/201936135}

\bibitem[{{Matsuno} {et~al.}(2019){Matsuno}, {Aoki}, \& {Suda}}]{matsuno19}
{Matsuno}, T., {Aoki}, W., \& {Suda}, T. 2019, \apjl, 874, L35,
  \dodoi{10.3847/2041-8213/ab0ec0}

\bibitem[{{McMillan}(2017)}]{mc17}
{McMillan}, P.~J. 2017, \mnras, 465, 76, \dodoi{10.1093/mnras/stw2759}

\bibitem[{{Myeong} {et~al.}(2018{\natexlab{a}}){Myeong}, {Evans}, {Belokurov},
  {Sand ers}, \& {Koposov}}]{myeong18b}
{Myeong}, G.~C., {Evans}, N.~W., {Belokurov}, V., {Sand ers}, J.~L., \&
  {Koposov}, S.~E. 2018{\natexlab{a}}, \apjl, 856, L26,
  \dodoi{10.3847/2041-8213/aab613}

\bibitem[{{Myeong} {et~al.}(2018{\natexlab{b}}){Myeong}, {Evans}, {Belokurov},
  {Sand ers}, \& {Koposov}}]{myeong18a}
---. 2018{\natexlab{b}}, \mnras, 478, 5449, \dodoi{10.1093/mnras/sty1403}

\bibitem[{{Myeong} {et~al.}(2019){Myeong}, {Vasiliev}, {Iorio}, {Evans}, \&
  {Belokurov}}]{myeong19}
{Myeong}, G.~C., {Vasiliev}, E., {Iorio}, G., {Evans}, N.~W., \& {Belokurov},
  V. 2019, \mnras, 488, 1235, \dodoi{10.1093/mnras/stz1770}

\bibitem[{{Naidu} {et~al.}(2020){Naidu}, {Conroy}, {Bonaca}, {Johnson}, {Ting},
  {Caldwell}, {Zaritsky}, \& {Cargile}}]{naidu20}
{Naidu}, R.~P., {Conroy}, C., {Bonaca}, A., {et~al.} 2020, arXiv e-prints,
  arXiv:2006.08625.
\newblock \doarXiv{2006.08625}

\bibitem[{{Ngeow} {et~al.}(2020){Ngeow}, {Belecki}, {Burruss}, {Drake},
  {Graham}, {Kaplan}, {Kupfer}, {Mahabal}, {Masci}, {Riddle}, {Rodriguez}, \&
  {Rusholme}}]{ngeow20}
{Ngeow}, C.-C., {Belecki}, J., {Burruss}, R., {et~al.} 2020, \aj, 160, 31,
  \dodoi{10.3847/1538-3881/ab930b}

\bibitem[{{Pawlowski} {et~al.}(2012){Pawlowski}, {Pflamm-Altenburg}, \&
  {Kroupa}}]{pawlowski12}
{Pawlowski}, M.~S., {Pflamm-Altenburg}, J., \& {Kroupa}, P. 2012, \mnras, 423,
  1109, \dodoi{10.1111/j.1365-2966.2012.20937.x}

\bibitem[{{Price-Whelan} \& {Bonaca}(2018)}]{price18}
{Price-Whelan}, A.~M., \& {Bonaca}, A. 2018, \apjl, 863, L20,
  \dodoi{10.3847/2041-8213/aad7b5}

\bibitem[{{Ramos} {et~al.}(2020){Ramos}, {Mateu}, {Antoja}, {Helmi},
  {Castro-Ginard}, {Balbinot}, \& {Carrasco}}]{ramos20}
{Ramos}, P., {Mateu}, C., {Antoja}, T., {et~al.} 2020, \aap, 638, A104,
  \dodoi{10.1051/0004-6361/202037819}

\bibitem[{{Riley} \& {Strigari}(2020)}]{riley20}
{Riley}, A.~H., \& {Strigari}, L.~E. 2020, \mnras,
  \dodoi{10.1093/mnras/staa710}

\bibitem[{{Simion} {et~al.}(2014){Simion}, {Belokurov}, {Irwin}, \&
  {Koposov}}]{simion14}
{Simion}, I.~T., {Belokurov}, V., {Irwin}, M., \& {Koposov}, S.~E. 2014,
  \mnras, 440, 161, \dodoi{10.1093/mnras/stu133}

\bibitem[{{Simion} {et~al.}(2019){Simion}, {Belokurov}, \&
  {Koposov}}]{simion19}
{Simion}, I.~T., {Belokurov}, V., \& {Koposov}, S.~E. 2019, \mnras, 482, 921,
  \dodoi{10.1093/mnras/sty2744}

\bibitem[{{Sohn} {et~al.}(2018){Sohn}, {Watkins}, {Fardal}, {van der Marel},
  {Deason}, {Besla}, \& {Bellini}}]{sohn18}
{Sohn}, S.~T., {Watkins}, L.~L., {Fardal}, M.~A., {et~al.} 2018, \apj, 862, 52,
  \dodoi{10.3847/1538-4357/aacd0b}

\bibitem[{{Vasiliev}(2019{\natexlab{a}})}]{eugene19}
{Vasiliev}, E. 2019{\natexlab{a}}, \mnras, 484, 2832,
  \dodoi{10.1093/mnras/stz171}

\bibitem[{{Vasiliev}(2019{\natexlab{b}})}]{agama}
---. 2019{\natexlab{b}}, \mnras, 482, 1525, \dodoi{10.1093/mnras/sty2672}

\bibitem[{{Xue} {et~al.}(2008){Xue}, {Rix}, {Zhao}, {Re Fiorentin}, {Naab},
  {Steinmetz}, {van den Bosch}, {Beers}, {Lee}, {Bell}, {Rockosi}, {Yanny},
  {Newberg}, {Wilhelm}, {Kang}, {Smith}, \& {Schneider}}]{xue08}
{Xue}, X.~X., {Rix}, H.~W., {Zhao}, G., {et~al.} 2008, \apj, 684, 1143,
  \dodoi{10.1086/589500}

\bibitem[{{Xue} {et~al.}(2014){Xue}, {Ma}, {Rix}, {Morrison}, {Harding},
  {Beers}, {Ivans}, {Jacobson}, {Johnson}, {Lee}, {Lucatello}, {Rockosi},
  {Sobeck}, {Yanny}, {Zhao}, \& {Allende Prieto}}]{xue14}
{Xue}, X.-X., {Ma}, Z., {Rix}, H.-W., {et~al.} 2014, \apj, 784, 170,
  \dodoi{10.1088/0004-637X/784/2/170}

\bibitem[{{Yanny} {et~al.}(2009){Yanny}, {Rockosi}, {Newberg}, {Knapp},
  {Adelman-McCarthy}, {Alcorn}, {Allam}, \& {Allende Prieto}}]{yanny09}
{Yanny}, B., {Rockosi}, C., {Newberg}, H.~J., {et~al.} 2009, \aj, 137, 4377,
  \dodoi{10.1088/0004-6256/137/5/4377}

\bibitem[{{Yuan} {et~al.}(2018){Yuan}, {Chang}, {Banerjee}, {Han}, {Kang}, \&
  {Smith}}]{yuan18}
{Yuan}, Z., {Chang}, J., {Banerjee}, P., {et~al.} 2018, \apj, 863, 26,
  \dodoi{10.3847/1538-4357/aacd0d}

\bibitem[{{Yuan} {et~al.}(2019){Yuan}, {Smith}, {Xue}, {Li}, {Liu}, {Wang},
  {Li}, \& {Chang}}]{yuan19}
{Yuan}, Z., {Smith}, M.~C., {Xue}, X.-X., {et~al.} 2019, \apj, 881, 164,
  \dodoi{10.3847/1538-4357/ab2e09}

\bibitem[{{Yuan} {et~al.}(2020){Yuan}, {Myeong}, {Beers}, {Evans}, {Lee},
  {Banerjee}, {Gudin}, {Hattori}, {Li}, {Matsuno}, {Placco}, {Smith},
  {Whitten}, \& {Zhao}}]{yuan20}
{Yuan}, Z., {Myeong}, G.~C., {Beers}, T.~C., {et~al.} 2020, \apj, 891, 39,
  \dodoi{10.3847/1538-4357/ab6ef7}

\bibitem[{{Zhao} {et~al.}(2012){Zhao}, {Zhao}, {Chu}, {Jing}, \&
  {Deng}}]{zhao12}
{Zhao}, G., {Zhao}, Y.-H., {Chu}, Y.-Q., {Jing}, Y.-P., \& {Deng}, L.-C. 2012,
  Research in Astronomy and Astrophysics, 12, 723,
  \dodoi{10.1088/1674-4527/12/7/002}

\end{thebibliography}
\bibliographystyle{aasjournal}

\end{document}